\documentclass[aps, nofootinbib]{revtex4}

\usepackage{amsmath}
\usepackage{amssymb}
\usepackage[figuresright]{rotating}
\usepackage{color}

\newcommand{\ket}[1]{| #1\rangle}
\newcommand{\bra}[1]{\langle #1 |}

\newcommand{\BQIC}{Berkeley Quantum Information and Computation Center, Berkeley, California 94720 USA}
\newcommand{\DeptChem}{Department of Chemistry, University of California, Berkeley, California 94720 USA}
\newcommand{\LBNL}{Physical Biosciences Division, Lawrence Berkeley National Laboratory, 
Berkeley, California 94720, USA}

\begin{document}

\title{Quantum entanglement phenomena in photosynthetic light harvesting complexes}

\author{K. Birgitta Whaley$^{1,2}$, Mohan Sarovar$^{1,2}$, Akihito Ishizaki$^{2,3}$}
\address{1. \BQIC \\ 2. \DeptChem \\ 3. \LBNL}

\begin{abstract}
We review recent theoretical calculations of quantum entanglement in photosynthetic light harvesting complexes. These works establish, for the first time, a manifestation of this characteristically quantum mechanical phenomenon in biologically functional structures. We begin by summarizing calculations on model biomolecular systems that aim to reveal non-trivial characteristics of quantum entanglement in non-equilibrium biological environments. We then discuss and compare several calculations performed recently of excitonic dynamics in the Fenna-Matthews-Olson light harvesting complex and of the entanglement present in this widely studied pigment-protein structure. We point out the commonalities between the derived results and also identify and explain the differences. We also discuss recent work that examines entanglement in the structurally more intricate light harvesting complex II (LHCII).
During this overview, we take the opportunity to clarify several subtle issues relating to entanglement in such biomolecular systems, including the role of entanglement in biological function, the complexity of dynamical modeling that is required to capture the salient features of entanglement in such biomolecular systems, and the relationship between entanglement and other quantum mechanical features that are observed and predicted in light harvesting complexes. Finally, we suggest possible extensions of the current work and also review the options for experimental confirmation of the predicted entanglement phenomena in light harvesting complexes.
\end{abstract}

\maketitle

\section{Introduction}
\label{sec:intro}

Entanglement, the non-local correlations between distinct quantum systems, was referred to by Schr\"odinger in his 1935 discussion of the issues raised by the paradox of Einstein, Podolsky and Rosen (EPR) \cite{Ein.Pod.etal-1935}, as ``the characteristic trait of quantum mechanics, the one that enforces its entire departure from classical lines of thought'' \cite{Sch-1935a}.  The phenomenon is uniquely quantum in that, unlike wave-like evolutions, there is no classical analogue.  The non-local correlations derive from the structure of the quantum states of the combined system.  The fact that these quantum correlations in the composite system are independent of the spatial separation of the constituents, so that in some sense these appear to behave as a single system, results in violation of local realism, leading to the difficulties posed by EPR  \cite{Ein.Pod.etal-1935}.  Entanglement may also be viewed from a knowledge perspective - an entangled pure state of the composite system, for which the von Neumann entropy is zero, corresponding to maximum knowledge, cannot be expressed in terms of states of maximal knowledge of its constituents.  As Schr\"odinger put it, ``the best possible knowledge of a whole does not necessarily include the best possible knowledge of its parts'' \cite{Sch-1935a}.  Analysis of entanglement and its consequences for measurement and the conceptual foundations of quantum mechanics led to development of statistical inequalities (e.g. Bell \cite{Bel-1964}, CHSH \cite{Cla.Hor.etal-1969}, Leggett-Garg \cite{Leg.Gar-1985}) for experimental verification of the `spooky action at a distance', as Einstein informally referred to it.  More recently, the growth of quantum information processing over the last 15 years has given rise to a large and increasing literature of entanglement theory, because of the key role that entangled states play in quantum communication and in enabling non-trivial quantum computation \cite{mikeandike}. Here, entanglement has emerged as a major resource for realizing quantum protocols.  An increasing number of applications of entanglement theory are also now being made to analysis of complex many-particle quantum systems and are  bringing new insight into the nature and role of many-particle correlations in these systems \cite{Ami.Faz.etal-2008}.  

The discussion of entanglement now extends far beyond its original context of bipartite pure states, encompassing also mixed states, multipartite states, combinations of these, as well as analysis of entanglement in large scale quantum many body systems.  While a complete understanding of entanglement in complex systems is still absent, it is evident that the non-entangled, separable states generally constitute only a very small subset of all states in the Hilbert space of a physical system. The development of measures to quantify entanglement in these various situations for both discrete and continuous quantum systems is an active area of theoretical research \cite{Ple.Vir-2007}. The difficulty of developing calculable measures of entanglement is worth bearing in mind when seeking to uncover quantum correlations in highly complex dynamical systems such as are encountered in biological settings.  

In this paper we review recent discussions of entanglement in photosynthetic light harvesting systems that have been prompted by experimental measurements of excitonic energy transport showing evidence of quantum coherence. Quantum coherences have now been observed in two-dimensional spectroscopic studies \cite{Cho-2009} of energy transfer within several different light harvesting complexes.  Experiments with the Fenna-Matthews-Olson (FMO) complex, a pigment-protein complex in green sulfur bacterium that transports electronic energy from peripheral light harvesting antennas to the reaction center, have shown the presence of quantum beats between excitonic levels at both cryogenic (77\,K) and ambient (300\,K) temperatures \cite{Eng.Cal.etal-2007,Pan.Hay.etal-2010}. Similar evidence for quantum coherence has now also been seen in light harvesting antenna complexes of green plants \cite{Cal.Gin.etal-2009} and marine algae \cite{Col.Won.etal-2010}. These experimental studies, coupled with related studies of quantum coherence within the reaction center \cite{Lee.Che.etal-2007}, now constitute a growing body of evidence for manifestation of quantum coherent effects in energy transfer within both plant and bacterial 
light harvesting complexes.  In all systems studied to date, electronic excitations of the relevant chromophores are closely coupled to vibrational modes of the surrounding protein scaffold.  This has generated a lot of interest in understanding how such coherence can persist within an open quantum system characterized by both static and dynamic disorder, dephasing and dissipative interactions with the protein environment.  This understanding is a prerequisite to determining whether entanglement may be found and how it is manifested.  Because of the close coupling of electronic and vibrational motions, analysis of the entanglement of the electronic excitations requires using measures of entanglement for mixed states, regardless of whether the analysis refers to {\it in vivo} or  laboratory conditions.  The studies reviewed below show that both bipartite and multipartite decompositions of the excitonic Hilbert space yield insight into the nature of quantum correlations in these biological systems.  
To set the scene, we shall first review studies of entanglement in generic 
chromophoric systems that mimic various aspects of a biological environment.  We shall then move to analysis of the entanglement studies that have been made for various theoretical models of the FMO complex and light harvesting complex II (LHCII).

\section{Entanglement in model systems}
\label{sec:model}

Dimeric models provide the smallest systems in which features of entanglement can be studied.  While energy transport is unrealistically restricted, analysis of excitonic dynamics in dimeric systems has revealed a number of insights into bipartite entanglement and how it may be affected by the biological environment.  The first study of entanglement in a dimer of chromophores was made by Thorwart et al., \cite{Tho.Eck.etal-2009}  who used a spin-boson Hamiltonian model for each chromophore together with dipole-dipole interchromophore coupling and an ohmic bath with variable frequency cutoff.  Using the real time path integral approach to simulate the excitonic dynamics, these authors evaluated the time dependence of the {\it negativity} measure of bipartite entanglement, $N(t) = \max \left[ 0, -2 \xi_{\rm min}(t) \right]$ ($\xi_{\rm min}$ is the smallest eigenvalue of the partial transpose reduced density operator of one chromophore) \cite{Vid.Wer-2002} and investigated the behavior of this as the bath frequency cut off, and hence the timescale of the bath response, varies. Starting from an initially entangled single exciton state, it was found that the negativity decays rapidly for a quasi-Markovian bath (high frequency cutoff) while the decay can be significantly delayed with a slow bath.  When a separable, non-entangled two-exciton initial state was used with a slow bath, the negativity showed an initial rise followed by an oscillatory decay, showing that bipartite entanglement can be generated and modulated by the interaction with a common bath.  A polaronic analysis showed that the long wave-length bath modes are most efficient in this generation of entanglement.    Dynamic entanglement modulation has also been identified as a generic consequence of driving a two-site model by classical harmonic oscillations by Briegel and co-workers \cite{Cai.Pop.etal-2008}.  These authors showed that a dimer of two coupled quantum spins in the presence of a Markovian bath (described by a temperature independent Lindblad equation) can not only evolve into an entangled state (measured in this study in terms of the {\it concurrence} \cite{Hil.Woo-1997}) when the dimer Hamiltonian is driven by a classical oscillator, but that this entanglement will also recur persistently, even at temperatures too high to support any static entanglement in the non-driven dimer.  While this result is obtained for an ideal harmonic driving of all terms in the quantum Hamiltonian and no evidence for such persistent controlled entanglement in biological systems is given, as discussed by Briegel and co-workers the fact that it can be found in an open driven system far from equilibrium raises the question of whether such entanglement might not only be present in live organisms, but also play a functional role \cite{Bri.Pop-2008, Cai.Pop.etal-2008}. Briegel and Popescu distinguish such ``live'' entanglement that is present only when a system is driven from both the intrinsic entanglement that is present in the basic constituents of biological systems, e.g., entanglement of bonding electrons in molecules, and any entanglement that may accompany a biological process but is incidental or ``dead'' because it does not play a functional role \cite{Bri.Pop-2008}.  Another aspect of environmental dynamics was explored by Hossein-Nejad and Scholes \cite{Hos.Sch-2010} in a recent analysis of the effects of an anticorrelated phonon bath coupling on dimer entanglement.  This situation may arise from coupling to an interchromophoric vibrational mode and allows an exact solution of the dimer dynamics.  
Hossein-Nejad and Scholes evaluated the dynamics in a phonon-dressed exciton basis that results from this special coupling symmetry, to find that in such a system longer lived coherences may be found for more weakly coupled dimers, a somewhat counterintuitive result in stark contrast with the situation for dimers with independent bath couplings \cite{Ish.Fle-2009}.

These model studies have explored a variety of forms of exciton-phonon coupling.  It is important to motivate such couplings physically, even within a model calculation.  Here we note that the spin-boson Hamiltonian, which is ubiquitous in condensed matter physics, should be used with care for pigment-protein complexes.  In particular, the standard $\sigma_z$ form of coupling employed in the spin-boson case predicts zero reorganization energy when applied to a single chromophore.  When applied to dimer states in the single exciton subspace it yields a physically correct reorganization energy but imposes anticorrelations in the bath couplings: these can only be justified when physically motivated, e.g., as in Ref. \cite{Ish.Cal.etal-2010}.

A study of excitation transport and entanglement in biologically-inspired model systems composed of more than two chromophores has been conducted by Scholak et al. \cite{Sch.Mel.etal-2010}. 
Making dynamical simulation on random configurational samples of $N=10$ chromophores in a spherical volume, they examine the dependence of transport efficiency and degree of entanglement on the number of sites that are entangled. Inter-chromophoric coupling is determined by separation distance -- mimicking dipole-dipole coupling -- and two models of dynamics are simulated: (i) purely Hamiltonian and (ii) Hamiltonian plus a pure dephasing (temperature-independent) Lindblad model. 
Their results 
appear to show that for the purely Hamiltonian model of dynamics (which is highly artificial in any biological context), the more entangled configurations have a larger probability of having highly efficient transport. When dephasing dynamics are included, both the entanglement and transport efficiency are decreased but the correlation between extent of entanglement and efficiency persists. 
From these simulations the authors conclude that ``efficient transport unambiguously requires strong entanglement" and further claim that this is evidence for a functional role for multipartite entanglement in biomolecular -- e.g., light harvesting  -- systems. There are several reasons to question these particular claims of a biological role for entanglement.  Firstly, the analysis of  \cite{Sch.Mel.etal-2010} is all in the single excitation subspace, where as discussed in detail in Sec. \ref{sec:fmo} below, it has so far not been possible to ascribe a role to entanglement that is distinct from the role of coherence. Hence it could equally be concluded that the enhanced efficiency is a result of coherence. Secondly, to draw conclusions about biomolecular systems from simulations of purely Hamiltonian dynamics is misleading because biomolecular systems will almost surely never be completely coherent -- it is essential to model the decoherent dynamics accurately to get an reliable picture. Finally, it should also be noted that the measures of entanglement and efficiency used in Ref. \cite{Sch.Mel.etal-2010} are maximal quantities -- i.e., efficiency is the maximum achieved population over a fixed evolution time period, and $n$-partite entanglement is the maximum achieved $n$-partite entanglement over the same evolution period. This is in contrast to most other studies on light harvesting complexes 
that examine instead the integrated efficiency (e.g., \cite{Reb.Moh.etal-2009, Ple.Hue-2008, Sar.Che.etal-2009}) or time dependent entanglement (e.g., \cite{Sar.Ish.etal-2009,Ish.Fle-2010}) or both (e.g., \cite{Car.Chi.etal-2010,Fas.Ola-2010}).  It is this focus on maximal rather than time dependent or integrated quantities that leads to the conclusion of Scholak et al. that completely coherent dynamics lead to larger efficiency than combined coherent and dephasing dynamics, a finding that contradicts previous studies which generally observe that dephasing can boost quantum transport efficiency in theoretical models emulating generic light harvesting systems -- e.g. \cite{Reb.Moh.etal-2009, Ple.Hue-2008, Sar.Che.etal-2009, Moh.Reb.etal-2008, Gaa.Bar-2004, Reb.Moh.etal-2009a}
\footnote{However, it is only fair to point out that these prior studies are also based on relatively crude model descriptions of the system-bath interactions. The true relationship between quantum coherence/entanglement and efficiency of photosynthetic light harvesting is thus not yet conclusive. Indeed Ref. \cite{Ish.Fle-2010} gives one example showing that strong quantum coherence/entanglement can accelerate electronic energy transfer in certain circumstances.}.
Under completely coherent dynamics an excitation wavepacket can propagate to all parts of a chromophoric network and hence have large population on any site for a short period of time, leading to a large maximal value of entanglement as measured by Scholak et al. However, this does not mean that there will be efficient trapping of the excitation: for this to happen the excitation should reside at the trapping site for significant periods of time (since the trapping mechanism acts on slower timescales than the transport mechanism).  The former maximum value of population is what Scholak et al. use as a measure of efficiency in Ref. \cite{Sch.Mel.etal-2010}, while the time integrated trapping probability, which is the physically relevant metric, is what is analyzed in other work.

\section{Entanglement in the Fenna-Matthews-Olson complex}
\label{sec:fmo}

\begin{figure}[h!]
   \begin{center}
   \begin{tabular}{c}
   \includegraphics[height=4.9cm]{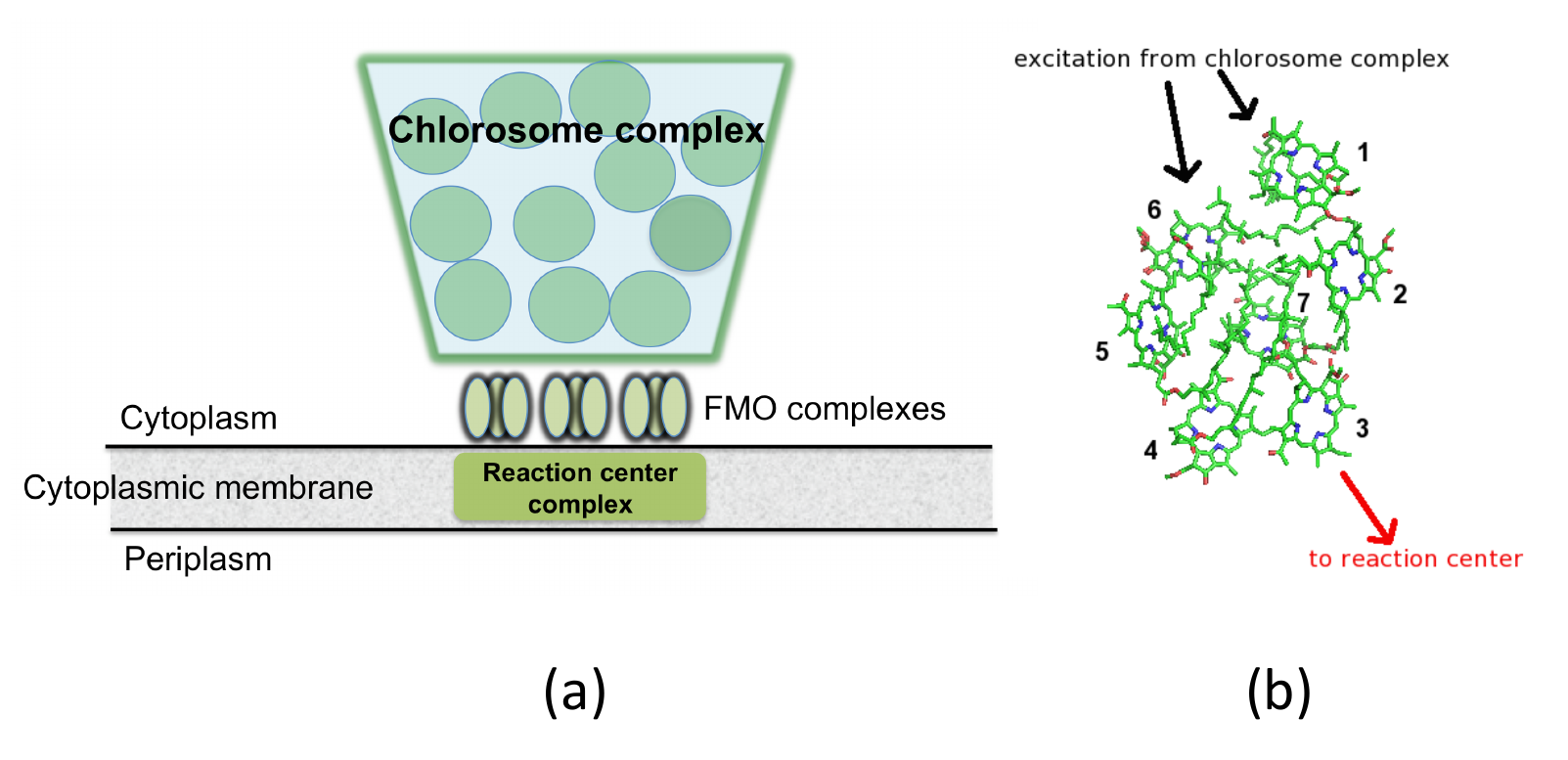}
   \end{tabular}
   \end{center}
   \caption[example] 
   { \label{fig:fmo} 
The Fenna-Matthews-Olson (FMO) complex is a widely studied pigment-protein light harvesting component
\cite{Fen.Mat-1975, Ame.Val-2000,Bla-2002}.
It is a critical unit of the light harvesting apparatus of green sulfur bacteria, where it functions as a conduit for excitation energy to pass from the primary light harvesting components -- the chlorosomes -- to the inter-membrane reaction center of these bacteria. Panel \textbf{(a)} shows the location of the FMO protein within the light harvesting apparatus of green sulfur bacteria. Panel \textbf{(b)} shows an FMO monomer (from the species \textit{Chlorobaculum tepidum}) with its protein scaffolding removed for clarity (Protein data bank ID: 3ENI). The green molecules are bacteriochlorophyll {\it a} (BChl{\it a}), and are labeled with a standard numbering order. Electronic excitations enter from the chlorosome complex through BChl{\it a}s 1 and 6, and exits to the reaction center primarily through BChl{\it a} 3.
}
   \end{figure}

Most theoretical studies of non-classical phenomena in light harvesting structures to date have focused on the FMO complex of green sulfur bacterium 
\cite{Fen.Mat-1975,Ame.Val-2000,Bla-2002} 
because this is an extremely well characterized pigment-protein structure.  
FMO does not directly absorb sunlight, but acts instead as a ``quantum wire'' to transmit electronic excitation from the light harvesting antenna that actually absorbs photons, to the reaction center where charge separation then occurs. 
Fig. \ref{fig:fmo} shows a schematic of the structural location and context of the FMO complex in the light harvesting apparatus, together with a detailed structure of the relative orientations of the seven molecular chromophores (bacteriochlorophyll {\it a}) in the complex.
The present authors have made a detailed study of the temporal duration and spatial extent of entanglement in the FMO complex using a simulation of excitation energy transfer (EET) dynamics under conditions that approximate the \textit{in vivo} environment \cite{Sar.Ish.etal-2009}. 
The theoretical approach based on a hierarchy of coupled master equations \cite{Ish.Fle-2009} employed for simulating EET dynamics in Ref. \cite{Sar.Ish.etal-2009} is able to describe the non-Markovian interplay between electronic excitations of the chromophores and their protein environment, including aspects such as site-dependent reorganization dynamics of the surrounding protein \cite{Gro.Yu.etal-1998,Ish.Cal.etal-2010}.
These play a significant role in photosynthetic EET since
the conditions for photosynthetic EET generally lie between the two extremes of weak and strong exciton-phonon coupling where conventional Redfield and F\"orster theories are respectively valid  \cite{Ish.Cal.etal-2010}.  The approach of Ref. \cite{Ish.Fle-2009} is valid for all coupling strengths, while reducing to these theories in their respective limits of validity: it thus provides the key capability to interpolate between these two limits that is necessary for reliable modeling of photosynthetic EET.  Such a high accuracy model is essential if one seeks a quantitative understanding of delicate quantum properties such as entanglement in these complex environments.  

\begin{figure}[h!]
   \begin{center}
   \begin{tabular}{c}
   \includegraphics[height=5cm]{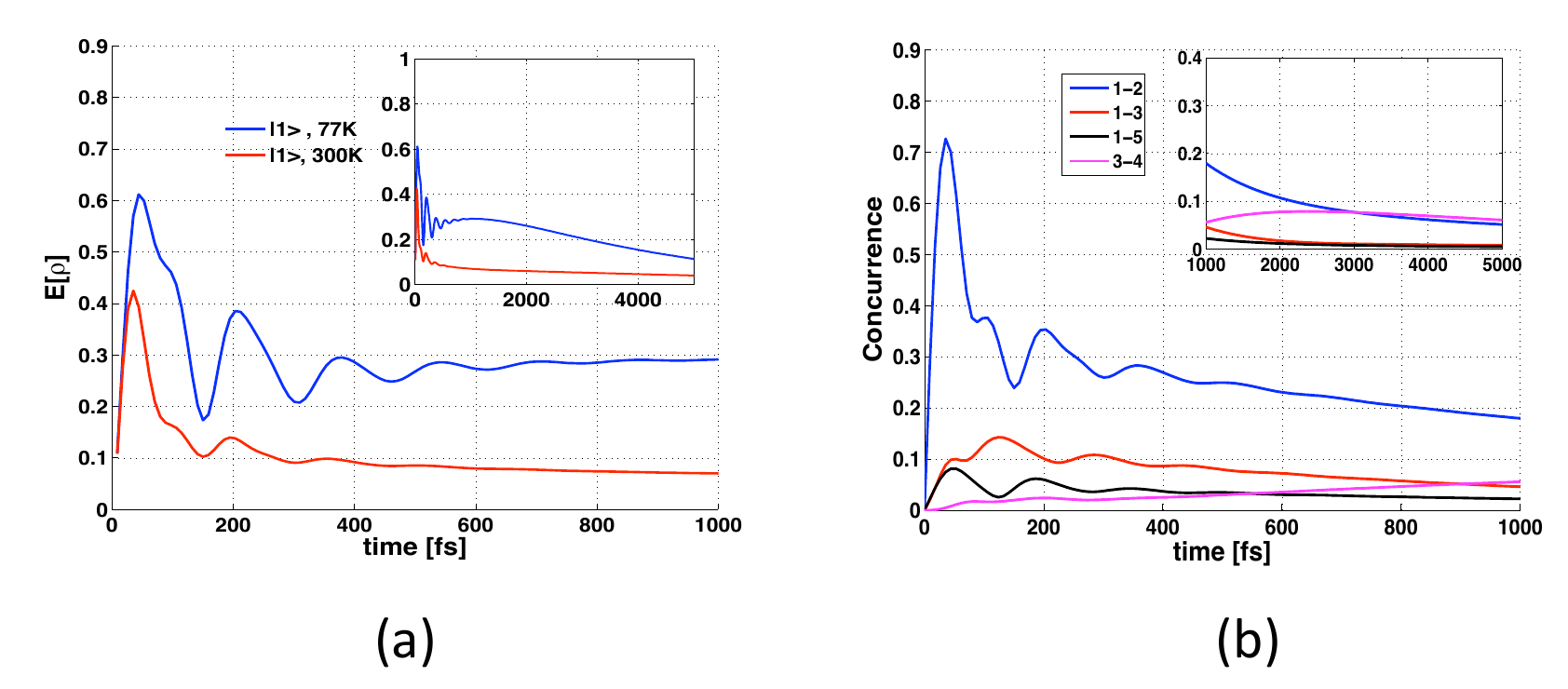}
   \end{tabular}
   \end{center}
   \caption[example] 
   { \label{fig:entangleFMO} 
Measures of entanglement in the FMO complex calculated from simulations of energy transfer under physically accurate conditions. Figure adapted from Ref. \cite{Sar.Ish.etal-2009}. Panel \textbf{(a)} shows the time evolution of the global entanglement measure after an initial localized excitation of site 1. The blue and red curves, which represent evolution at T=77K and T=300K, respectively, show that entanglement in the FMO complex remains finite for long times (the inset shows time evolution up to 5ps). Panel \textbf{(b)} shows the concurrence measure of bipartite entanglement between various BChl{\it a} molecules as a function of time following initial excitation on BChl{\it a} 1 at time $t=0$, evaluated at temperature 300K. The inset in this panel shows the long-time behavior (up to 5ps).  The figure clearly shows the presence of long-time entanglement between the ``distant" chromophores 1 and 3.
}
   \end{figure}

In Ref. \cite{Sar.Ish.etal-2009}, Sarovar et al. study entanglement in the ``site basis" that corresponds to non-classical correlations of the electronic excited states of distinct and spatially separated pigments in the FMO complex. For molecular structures such as the FMO complex that receive a very small influx of excitation energy, \footnote{It should be noted that in higher plants, approximately 10 photons are absorbed per chlorophyll per second 
under conditions of full sunlight
\cite{Bla-2002}.
In contrast, green sulfur bacteria are usually found below the surface of 
lakes, oceans and similar environments because they are obligately anaerobic. Consequently, the number of sunlight photons 
absorbed per bacteriochlorophyll in the FMO complex is much smaller. \label{foot-FMO}}
only the zero and single excitation subspaces are relevant to EET under {\it in vivo} conditions.  
Sarovar et al. show that in these low lying subspaces, entanglement and coherence are inextricably related, with each being necessary and sufficient for the other \cite{Sar.Ish.etal-2009}. 
 Consequently a witness for entanglement can also serve as a witness for coherence, and vice versa. 
 In order to assess overall trends in the entanglement 
lifetime, a \textit{global measure of entanglement} that is appropriate to the conditions typical for light harvesting complexes such as the FMO complex was derived. This global measure, $E[\rho]$, where $\rho$ is the electronic (pigment) density matrix, is based on the relative entropy of entanglement \cite{Ved.Ple.etal-1997} and can be shown to be an entanglement monotone in the restricted Hilbert space $\mathcal{H}_0 \oplus \mathcal{H}_1$, where $\mathcal{H}_i$ is the Hilbert space of states with $i$ excitations, under the set of local operations accessible during EET under the {\it in vivo} conditions \cite{Sar.Ish.etal-2009}. Essentially this is the relative entropy of entanglement restricted under superselection rules \cite{Wis.Bar.etal-2004,Bar.Rud-2007} relevant to energy transfer \textit{in vivo}. 
 While the numerical value of $E[\rho]$ (bounded by ln7 for FMO) does not appear have a operational interpretation as far as we know today, its utility derives from its ability to capture the presence or absence of entanglement over the entire complex and to assess the extent to which this saturates.
 Fig.~\ref{fig:entangleFMO} shows the behavior of this global measure of entanglement together with the \textit{concurrence}, a measure of bipartite entanglement \cite{Hil.Woo-1997} that captures the non-classical correlations between any two pigments in the complex for an initial excitation localized on chromophore 1.
 Evaluation of $E[\rho]$ for the physically relevant initial conditions, reveals that global entanglement over the entire FMO complex is actually generated by the pigment-protein dynamics, with an initial rise within 100 fs to $\sim 30-40 \%$ of the maximum possible value, that is then followed by strong oscillations which damp out over a time of several hundred fs to a long time contribution on the ps timescale (up to $\sim 5\,\textrm{ps}$). This long time persistent entanglement can be as much as $15 \%$ of the maximum and appears to be limited only by the relatively slow trapping dynamics of the reaction center, which were characterized by a rate of 4 ps$^{-1}$ in these simulations. 
Comparison of the global entanglement with the concurrence between specific pairs of chromophores showed that the entanglement in the FMO complex is true multipartite entanglement between at least $4-5$ pigments. Most significantly, the concurrence measure revealed that significant entanglement is generated and maintained between pigments 1 and 3, which are separated by $\sim 30 \,\AA$ -- the second largest separation distance in the FMO complex. At this distance there is negligible direct excitonic interaction between these two pigments; they are nevertheless brought into an entangled state via the mediating interactions with neighboring pigments.  Finally, the realistic simulation approach of Ref. \cite{Sar.Ish.etal-2009} allowed analysis of the temperature dependence of entanglement.  The results show that in FMO this is relatively robust to temperature effects, showing a decrease by only $\sim 25\%$ when the temperature is increased almost $4$-fold, from 77\,K to 300\,K.
 
As noted above, an accurate and realistic analysis of the excitonic dynamics is essential for revealing the nature and extent of entanglement in natural light harvesting systems.  The analysis in Ref.~\cite{Sar.Ish.etal-2009} was made to mimic {\it in vivo} conditions where FMO receives single excitations from the actual antenna system via a baseplate complex, rather than simulating the non-linear spectroscopy experiments
\footnote{In the experimental context, this corresponds to following the temporal evolution of entanglement only during what would correspond to the population time of a 2D photon echo experiment (with the first two pulses designed to initialize the excited state population at a single site), i.e., not incorporating the non-natural influence of laser pulses.}.
To analyze the effect of using less realistic dynamic models of light harvesting systems, Ref.~\cite{Sar.Ish.etal-2009} compared the time dependence of the predicted {\it in vivo} entanglement from the above realistic non-Markovian simulation of FMO dynamics that includes simulation of the natural time scale bath dynamics, with the corresponding results obtained from a Markovian description within Redfield theory \cite{Nit-2006}. The entanglement was calculated for dynamical evolution under both the full Redfield equations and the secular approximation to these, which is in the Lindblad form for Markovian dynamics, using the same bath spectral density as the realistic simulations. The Markovian simulations also yield entanglement persisting to ps timescales, but with significantly less oscillation, in both full Redfield and Lindblad descriptions.  Both of these also give quantitatively inaccurate values of $E[\rho]$ at short times, with the Lindblad description considerably underestimating and the full Redfield description significantly overestimating the amount of entanglement generated in the first 100 fs. 

Since entanglement is an enabling feature for quantum information processing \cite{Joz.Lin-2003}, an intriguing question that was already raised when the first experimental indications of long-lived electronic quantum coherences were found~\cite{Eng.Cal.etal-2007}, is whether these exclusively quantum correlations are associated with some form of quantum speedup analogous to that found in quantum algorithms.  This was examined in a detailed study by Hoyer et al. \cite{Hoy.Sar.etal-2009}, who mapped the dominant coupling pathways in the three-dimensional FMO to a one-dimensional graph and then evaluated the rate of spreading of an initial excitation on this graph in the presence of coupling to the non-Markovian protein bath, using the realistic simulation results of Ref.~\cite{Sar.Ish.etal-2009}.  This study shows that the quadratic speedup of spreading associated with a quantum random walk on an ideal graph characterized by equal site energies and absence of disorder, is found only at very short times, less than $\sim$~100 fs, which is the timescale corresponding to coherent transfer between neighboring strongly coupled chromophores.  At longer times the excitation spreading becomes linear or sublinear in time, corresponding to a diffusive or subdiffusive dynamical regime with no associated quantum speedup.  This analysis, the results of which were found to be independent of the specific details  of the underlying full-dimensional simulation, leads to the conclusion that there is no quantum speedup that would characterize the realization of a quantum algorithm in these systems.  In particular, while the FMO complex acts as a quantum wire to transmit excitation energy with extremely high quantum efficiency and manifests both quantum coherence and entanglement, it does not perform the analog of a quantum search in an open quantum system.   This key study indicated that if there is a biological role for the coherence and entanglement, it is not that of enabling natural quantum information processing.

The extent of entanglement possible within Markovian descriptions of the FMO complex has also been analyzed 
by Caruso et al. \cite{Car.Chi.etal-2009,Car.Chi.etal-2010} using a realistic model of the FMO pigment electronic Hamiltonian together with idealized models for the protein bath and the pigment coupling to this. These authors used the \textit{logarithmic negativity} bipartite measure of entanglement \cite{Vid.Wer-2002} to quantify the electronic entanglement between various bipartite partitions of the seven pigments.  Ref. \cite{Car.Chi.etal-2009} reports results of calculations within the single excitation subspace relevant to FMO {\it in vivo}, with a temperature-independent Markovian Lindblad description and artificially (i.e., non-natural) optimized environmental coupling rates.  While this does not purport to be a realistic representation of the effects of the protein environment, the calculations interestingly do show the initial rise and large oscillations seen in the more realistic study of Ref. \cite{Sar.Ish.etal-2009}, although the entanglement appears to subsequently decay unrealistically fast in \cite{Car.Chi.etal-2009}.
This short-lived nature is a result of using a temperature-independent Markovian Lindblad equation 
for the dynamics, which artificially 
activates the system toward an infinite temperature and thus destroys coherent features more rapidly. 
In Ref. \cite{Car.Chi.etal-2010} these calculations are extended to non-Markovian environmental fluctuation dynamics, by coupling all pigments to a common vibrational mode that is damped by coupling to a zero-temperature bath.  This extended model shows that temporal correlations in the bath can lead to longer lived excitonic entanglement. Ref.  \cite{Car.Chi.etal-2010}  also considers formulation of entanglement in the FMO complex in the ``exciton basis" -- i.e., the basis that diagonalizes the electronic Hamiltonian of the FMO complex. While entanglement in this basis has no clear interpretation in terms of non-classical correlation of distinct physical entities, the authors show that it has a similar timescale for decay 
(under the same temperature-independent Markovian model) as entanglement in the site basis.  

These studies, like those in Ref. \cite{Sar.Ish.etal-2009}, examine entanglement in the FMO complex conditioned on the fact that an excitation was absorbed by the FMO complex from the chlorosome baseplate (see Fig. \ref{fig:fmo}) at $t=0$. This formulation corresponds implicitly to analysis of a single FMO complex and is justifiable by the low excitation rate of FMO {\it in vivo} (see footnote \ref{foot-FMO}).  
Caruso et al.  also investigate the entanglement under thermal excitation of the FMO complex, corresponding implicitly to the injection of a thermal distribution of photons from the baseplate, which results in an indeterminate excitation time \cite{Car.Chi.etal-2010}.  The resulting statistically averaged entanglement is understandably considerably less than the entanglement resulting from a well-defined initial excitation time. These calculations were also used to investigate the 
effect of spatial correlations in the dephasing noise on the pigments.  The results indicated that such spatial correlations can enhance entanglement while also decreasing the excitation transport efficiency, consistent with other recent studies of the effects of spatial correlations on coherence and energy transport efficiency 
\cite{Reb.Moh.etal-2009a, Hen.Bel.etal-2009, Fas.Naz.etal-2009, Sar.Che.etal-2009, Wu.Liu.etal-2010, Moh.Sha.etal-2010}.
Finally, Ref.  \cite{Car.Chi.etal-2010}  also explores the entanglement generation during a laser excitation pulse, which introduces the possibility of multi-photon excitation, although the relative contributions of one- and two-exciton entanglement were not reported, and moreover multi-exciton physics (e.g., scattering and annihilation of excitons) was not simulated. 

Another recent study of non-classical correlations in the FMO complex quantifies not only entanglement but also \textit{quantum discord} \cite{Oll.Zur-2001} during dynamical evolution following an excitation of pigment 1 or 6 \cite{Bra.Wil.etal-2009}. Quantum discord is a measure of a larger class of quantum correlations than entanglement and captures a larger set of non-classical states (it has also recently been given operational interpretations \cite{Cav.Aol.etal-2010,Mad.Dat-2010}). In Ref. \cite{Bra.Wil.etal-2009}, Bradler et al. show that under a
temperature-independent Markovian Lindblad approximation of FMO dynamics, the quantum discord between various bipartitions of the 7 pigments can be non-zero for appreciably longer times (e.g. $\sim 10-20\,{\rm ps}$ at 77\,K) than the entanglement.  Furthermore, they show that quantum correlations constitute a significant component  of the total correlations (quantum and classical) between certain bipartitions of the 7 pigments for short times ($\lesssim 1\,{\rm ps}$). The authors also perform a restricted numerical minimization over the combined single and double excitation subspaces to evaluate the relative entropy of entanglement of the state of the FMO electronic degrees of freedom and show that this quantity can be less than the corresponding quantity 
evaluated in the single excitation subspace (a quantity that is more easily calculated and that, as mentioned above, is an entanglement monotone only in the restricted Hilbert space $\mathcal{H}_0 \oplus \mathcal{H}_1$ \cite{Sar.Ish.etal-2009}).

A third recent study, due to Fassioli and Olaya-Castro, focuses on the spatial distribution of entanglement during dynamical evolution of the FMO complex under Markovian assumptions \cite{Fas.Ola-2010}. These authors employ a variant of the concurrence measure of entanglement to analyze the amount of entanglement produced before absorption by the reaction center -- which they term the \textit{entanglement yield} -- between various dissections of the 7 pigments in the FMO complex under different initial conditions (excitation starting on site 1 or site 6) and variable environmental conditions.  The latter include changes in reorganization energy and bath spatial correlation of a secular Redfield dynamical model, as well as changes in dephasing rate of a temperature-independent Markovian bath model.  This study shows that within both secular Redfield and temperature-independent Markovian descriptions, the presence of entanglement correlates inversely with efficient transport across FMO.
We note here that calculations of the quantum transport efficiency are also strongly dependent on the detailed model assumed for the coupling to the protein environment and its dynamics.  For example, the efficiency obtained within pure dephasing models can be very different from that obtained when relaxation is included, giving rise to quite different temperature dependence \cite{Ish.Fle-2009a}.  Most importantly for the correlation between entanglement and energy transport efficiency, within the realistic temperature dependent description provided by
the approach of Ref. \cite{Ish.Fle-2009},  it is shown in Ref.~\cite{Sar.Ish.etal-2009} that increasing the temperature lowers both the amount of entanglement and the excitation trapping efficiency. 

A key difference between Ref. \cite{Fas.Ola-2010} and the aforementioned studies of entanglement in the FMO complex is that while the latter do not ascribe any functional role to entanglement, Fassioli and Olaya-Castro make an explicit conjecture as to such a role, based on 
the correlation seen in their calculations for temperature-independent Markovian models and on the observation common to all calculations to date that the amount of entanglement produced can vary with the initial state.  These authors suggest that entanglement therefore provides a mechanism for controllably interfering the two different energy transfer pathways in the FMO complex and thereby modulating the efficiency of energy transfer.  
However, it should be noted that since the analysis in Ref. \cite{Fas.Ola-2010} pertains to the single excitation subspace, where any measure of entanglement is also a measure of coherence \cite{Sar.Ish.etal-2009}, it is not possible to separate the two phenomena and to ascribe a functional role to one versus the other, based on the quantities measured in Ref. \cite{Fas.Ola-2010}.  In particular, 
these calculations provide no evidence for a functional role of the non-classical correlations intrinsic to entanglement, which places a more stringent demand than the search for a functional role of quantum coherence.   

In summary, studies of excitation dynamics in the FMO complex in the single excitation subspace using various levels of theoretical simulation have revealed the presence of long-lived, multipartite entanglement representing non-classical correlations of the location of the excitation on separated pigments. Regardless of the particular measure used, the entanglement is seen to persist for $\sim 1-2\,{\rm ps}$ under perturbative and Markovian dynamical models and for over $4\,{\rm ps}$ under more accurate non-perturbative and non-Markovian models of excitation dynamics. It should be clarified that this is considered to be ``long-lived'' entanglement because the timescale of energy transfer in the FMO complex is on the order of a few picoseconds, and hence the entanglement persists for \textit{functional} timescales. Nevertheless, despite this convincing evidence for entanglement in a prototypical light harvesting complex, it is inconclusive at the present time whether the phenomenon plays any functional role in light harvesting. Thus it is not yet clear whether this long lived entanglement of electronic excitation should be regarded as `dead' or `live' entanglement in the sense of Ref. \cite{Bri.Pop-2008}.

\section{Entanglement in larger light harvesting complexes}
\label{sec:larger}

\begin{figure}
	\begin{center}
		\includegraphics[width=0.6\linewidth]{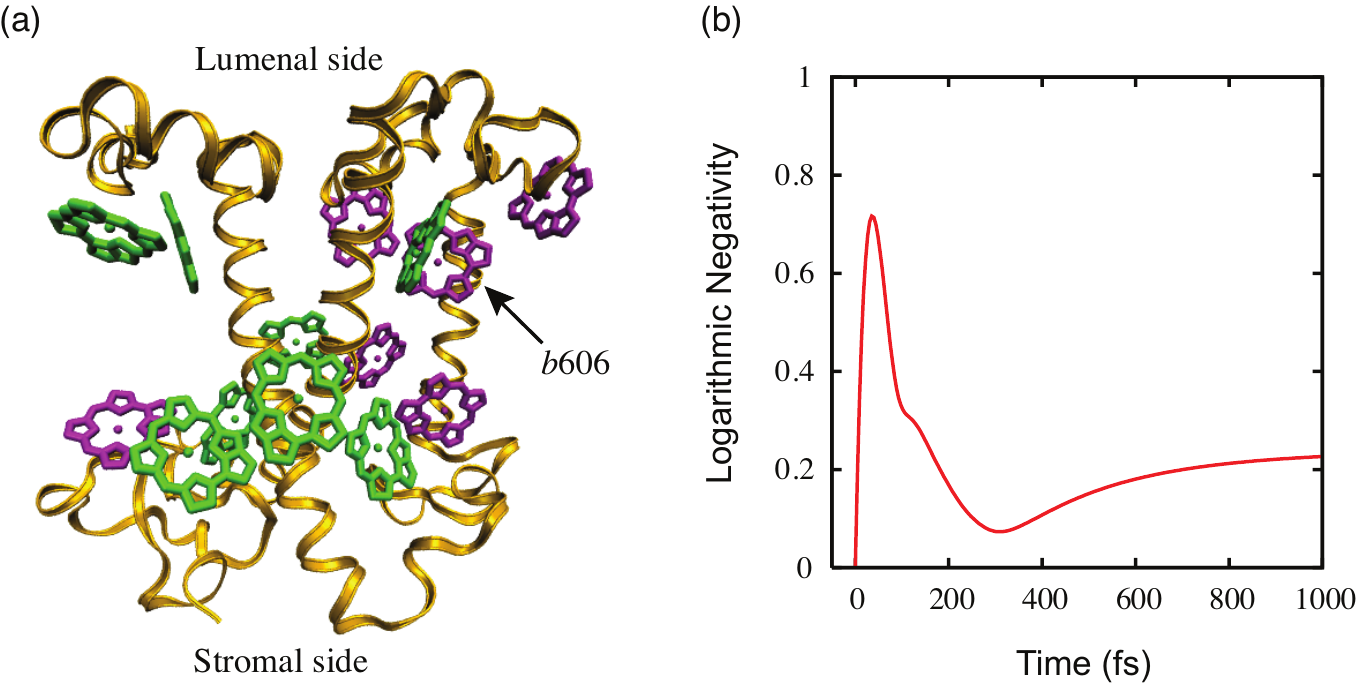}
	\end{center}
	\caption{
	{\bf (a)}
	The structural model of light-harvesting complex II (LHCII) of green plants obtained from X-ray crystallography (Protein data bank ID: 1RWT). 
	This panel shows the relative positions of the chlorophylls (Chl{\it a}
in green, Chl{\it b} in magenta) in a monomeric unit with the $\alpha$-helices spanning the membrane. 
	{\bf (b)}
Time evolution of bipartite quantum entanglement in LHCII as quantified by the logarithmic negativity. 
	Calculations were done for physiological temperature of 300\,K. \cite{Ish.Fle-2010}
	The bipartition is \{all the Chl{\it a}s\} $+$ \{all the  Chl{\it b}s\} and the initially excited pigment is Chl{\it b} 606. \cite{Ish.Fle-2010:note}
	}
	\label{LHCII-fig}
\end{figure}

Recently, it was demonstrated that light harvesting complex II (LHCII) also exhibits long-lived electronic coherence \cite{Cal.Gin.etal-2009}.
LHCII is the most abundant photosynthetic antenna complex in plants, containing over
50\% of the world's chlorophyll molecules \cite{Ame.Val-2000, Bla-2002}. 
The complex is a trimeric system composed of three monomeric subunits, each of which contains 8 chlorophyll {\it a} (Chl{\it a}) molecules and 6 chlorophyll {\it b} (Chl{\it b}) molecules, as depicted in Fig. \ref{LHCII-fig}(a).
Ishizaki and Fleming \cite{Ish.Fle-2010} have explored quantum entanglement in LHCII using the logarithmic negativity \cite{Vid.Wer-2002} to analyze entanglement across different bipartitions of the 14 chlorophyll pigments. Because the intensity of sunlight is weak, simultaneous excitation of multiple chlorophylls can be assumed to be almost impossible under physiological situations, even though the ratio of photons absorbed per chlorophyll is higher than in FMO
(see footnote \ref{foot-FMO} in Section \ref{sec:fmo}).
Consequently, as for FMO in Ref. \cite{Sar.Ish.etal-2009}, the initial condition for LHCII was also set to excitation of a single chlorophyll, in order to mimic conditions {\it in vivo}.
Figure \ref{LHCII-fig}(b) presents the entanglement dynamics at the physiological temperature of $300\,\mathrm{K}$ for the 
bipartition between the two different kinds of chlorophyll molecules, i.e., $A=$ \{all the Chl{\it a}s\} and $B=$ \{all the Chl{\it b}s\}, when the initially excited chlorophyll is Chl{\it b} 606. The plot shows that the excitation strongly entangles chromophores from partitions $A$ and $B$ at short times, after which the bipartite entanglement between the two types of chlorophylls decreases. However, at longer times there is a subsequent rise in entanglement to a persistent finite value 
which is a significant fraction of the maximum value.  This is a result of the steady state for the dynamical evolution being an exciton that is coherently delocalized across the bipartition (in this particular calculation no trapping, or inter-complex transfer dynamics was included -- both of these phenomena occur at significantly longer timescales than that shown in Fig. \ref{LHCII-fig}(b)). 

The plot in Fig. \ref{LHCII-fig}(b) show finite values of entanglement in the long time region, rather than the decay to zero seen in the Markovian and modified Markovian calculations for FMO \cite{Car.Chi.etal-2010,Fas.Ola-2010,Car.Chi.etal-2009}
As discussed above, such robust long-time entanglement is also a primary feature of the realistic studies of the FMO complex in Ref. \cite{Sar.Ish.etal-2009}. To shed light on the mechanism behind this phenomenon, which is not reproduced by the studies in Refs. \cite{Car.Chi.etal-2010,Fas.Ola-2010,Car.Chi.etal-2009}, we consider here the Hamiltonian of a pigment-protein complex.
Because neither nonadiabatic transitions nor radiative/nonradiative decays from the excited state  $\ket{\varphi_{m{\rm e}} }$ to the ground state $\ket{ \varphi_{m{\rm g}} }$ in the $m$th pigment are assumed in the dynamical simulations of Refs. \cite{Sar.Ish.etal-2009,Ish.Fle-2010}, i.e., only the dephasing interaction with the chromophoric environment is taken into account, the number of initial elementary excitations is conserved and thus the Hilbert space can be decomposed as
${\cal H}={\cal H}_0 \oplus {\cal H}_1 \oplus \cdots$,
where 
${\cal H}_n$
($n=0,1,\dots$) describes an $n$-exciton subspace comprising $n$ elementary excitations.
Therefore, Fig. \ref{LHCII-fig}(b) can be discussed in terms of the single-exciton subspace ${\cal H}_1$ alone, where the presence of a single excitation localized on the $m$th chlorophyll is expressed as $\ket{m} \equiv \ket{{\varphi_{m{\rm e}}}}\prod_{k\ne m}\ket{ \varphi_{k{\rm g}}}$.
The logarithmic negativity for this subspace \cite{Car.Chi.etal-2010} reduces then to
\begin{equation}
	E(A \vert B)
	=
	\log_2 
	\left(
		1+2\sqrt{\sum_{m\in A} \sum_{n\in B} | \bra{m} \rho \ket{n} |^2}.
	\right),
	\label{log-negativity-1exciton}
\end{equation}
Now $\ket{m}$ is not an eigenstate of the electronic Hamiltonian, because of the presence of the excitonic couplings $J_{mn}$.
Therefore, off-diagonal elements of the density matrices in the $\{\ket{m}\}$ basis do not necessarily vanish and it is then physically reasonable that the logarithmic negativity shows finite values in the long time region, i.e., shows persistent, robust quantum entanglement.

This analysis shows that the persistent entanglement in light harvesting complexes such as the FMO and LHCII complexes is physically reasonable even at ambient temperature, since in the physiologically relevant single-excitation subspace it is intimately related to the quantum coherence and delocalization of the excitons.  We see that, in contrast to these results from realistic simulations of the light harvesting EET, temperature-independent (or correspondingly infinite-temperature) theories such as the simplified Lindblad equation \cite{Bre.Pet-2002} or Haken-Strobl model \cite{Hak.Str-1973}, which oversimplify the treatment of the environment-induced fluctuation-dissipation processes, are not capable of producing this non-vanishing quantum entanglement in the equilibrium states \cite{Car.Chi.etal-2010,Fas.Ola-2010,Car.Chi.etal-2009}.
In Ref. \cite{Ish.Fle-2010}  the concurrence was also used to make a systematic analysis of the effect of the surrounding environment on the entanglement.  In particular, the influence of temperature, system-bath coupling strength, bath-relaxation timescales, and spatial correlation of fluctuations were studied. One interesting example in Ref. \cite{Ish.Fle-2010} shows that in certain physical circumstances spatial correlation of fluctuations can enhance quantum coherence/entanglement and also accelerate electronic energy transfer, in contrast with the more usual finding that enhancement of coherence correlates with a decrease in energy transfer \cite{Reb.Moh.etal-2009, Ple.Hue-2008, Sar.Che.etal-2009, Moh.Reb.etal-2008, Gaa.Bar-2004, Reb.Moh.etal-2009a} (see also footnote 1).

\section{Discussion}
\label{sec:disc}

The theoretical studies summarized here present convincing evidence for creation and robust evolution of quantum correlations in the natural functioning of pigment-protein structures such as light harvesting complexes.  The quantum non-local correlations associated with entanglement of distinct physical entities are seen to be naturally generated by the time evolution of excitonic energy through light harvesting structures and to be present both over relatively large distances of several nm and to be globally distributed over multiple chromophoric units.   Furthermore, the  studies of the FMO and LHCII light harvesting systems made using realistic dynamical simulations incorporating the environmental dynamics on their natural timescales \cite{Sar.Ish.etal-2009,Ish.Fle-2010} show that the entanglement is remarkably robust and persistent, achieving a significant long time, steady state value in both instances.  Another general feature that emerges from these more realistic studies is that while, as one might intuitively expect, quantum entanglement is suppressed with increasing temperatures, the entanglement is nevertheless still robust against environmental noise at ambient temperatures, continuing to show a significant steady state value on ps time scales at physiological temperatures.  This corresponds to the functional timescale of 
energy transport in light harvesting, raising intriguing questions as to whether and how this entanglement might be playing a biological role.  
While the studies summarized here show that the theoretical response to the question of whether \textit{in vivo} entanglement can be found is a resounding ''yes'', the functional relevance of this predicted entanglement in a biological system is still an open issue.  Theory has shown explicitly that despite these quantum correlations in the excitation energy transfer across a light harvesting complex, there is no quantum speedup in the quantum information processing sense associated with this highly efficient quantum process \cite{Hoy.Sar.etal-2009}. Indeed, not only is there no unambiguous demonstration of a unique role for entanglement as opposed to coherence for the {\it in vivo} conditions of light harvesting, there is also currently no clear understanding of the extent to which coherence plays a `live' functional role in photosynthesis. 

We can identify three key elements 
that enable the persistent quantum coherence and entanglement of electronic excitations in these structures. 
(i) The presence of moderate to strong excitonic coupling between constituent subsystems (molecular chromophores in most instances and certainly in the examples addressed here). 
(ii) A small to moderate coupling of the chromophores to their environment (as measured by, e.g., a small to moderate value of the resulting site reorganization energy), enabled by the embedding of light harvesting structures within membranes which shield them from the influence of solvent dynamics. 
(iii) The presence of complex protein environments, the dynamics of which are characterized by some degree of temporal and spatial correlation. 
These structural and environmental properties are crucial for the long-lived entanglement in light harvesting complexes and account for its persistence on functional timescales. 
As we have discussed here, to understand and accurately characterize the quantum correlations, it is essential to make a dynamical simulation with non-Markovian environmental dynamics coupled to the molecular chromophores with realistic couplings.  Markovian descriptions, while computationally attractive, do not show the characteristic persistent entanglement on functional timescales.  

Since the photosynthetic photon flux density is weak \cite{Bla-2002}, the single-excitation subspace is usually assumed to be of primary importance under physiological conditions. Hence the majority of the entanglement studies summarized here have focused on the analysis of dynamics in this subspace. The entanglement in this subspace is akin to \textit{mode entanglement} in quantum optics, illustrated for example by the entanglement of a single photon when distributed over multiple modes of the radiation field \cite{Enk-2005}.   Physically it is also manifested as coherent delocalization of the single particle's wave function. Such entangled states constitute a small subset of all the possible entangled states in the full Hilbert space describing the electronic states of $n$ chromophores.   For a dimer system, they do not span the full set of four Bell-like states, allowing only $\alpha|01\rangle \pm \beta|10\rangle$: the other two Bell-like states $\alpha|00\rangle \pm \beta|11\rangle$ require access to the two-excitation subspace.  The dynamics and lifetimes of more general entangled states in such systems are still largely unexplored.  While more general entangled states that have multiexciton components may not be of primary importance when considering physiological situations, they are nevertheless relevant for analysis of laboratory experiments. Nonlinear spectroscopic techniques such as two-dimensional electronic spectroscopy and photon echo measurements populate some higher excitation subspaces, e.g., the double-excitation subspace.  Recently, two-dimensional electronic double-quantum coherence spectroscopy has been proposed as a direct probe of double-excitation features of photosynthetic pigment-protein complexes such as the FMO complex \cite{Abr.Vor.etal-2008, Kim.Muk.etal-2009, Muk-2010}. Mukamel has suggested that entanglement of multiple excitations may be quite generally accessed by making appropriate correlation of  the time evolutions during selected control and delay periods during a multiple quantum coherence experiment \cite{Muk-2010}.
We point out that studies of entanglement in such multiexcitation systems will be complicated by at least two factors. Firstly, the dynamics of multiple excitations in a light harvesting complex is quite involved; exciton-exciton scattering and annihilation processes must be accounted for in order to capture the dynamics accurately \cite{Ame.Val-2000}, which considerably complicates the theoretical description beyond what has been attempted thus far. Secondly, the structure of entangled states becomes highly complex as more excitations are considered, and measures of entanglement become harder to formulate, although we note that measures across \textit{bi}partitions of the whole system do exist and can always be used to analyze the distribution of entanglement between specific bipartitions, e.g., as in \cite{Car.Chi.etal-2010, Ish.Fle-2010}. 

Although the biological significance of entanglement in light harvesting and more generally in photosynthesis is not yet clear, viewed from a different perspective, these entanglement studies have nevertheless brought new insight into the mechanism of excitation energy transfer in this important natural system.  In particular, we have seen that quantitative measures of entanglement such as those used here are useful tools in providing us with more detailed information regarding the interplay between the quantum delocalization and the surrounding environment than may be possible to learn by other means.  Thus the analysis of LHCII (Section \ref{sec:larger} and Ref. \cite{Ish.Fle-2010}) revealed dynamic localization and correlated fluctuation effects that are difficult to capture with a more traditional measure such as the inverse participation ratio but that are well assessed by the entanglement measures.
We note here that the role of the photosynthetic apparatus is not limited to efficiently transporting electronic excitation energy toward reaction center complexes. Especially at high light intensities, many light harvesting complexes have regulation mechanisms that initiate quenching of excess excitation energy in order to mitigate oxidative damage and protect reaction centers \cite{Bla-2002,Dem.Ada.etal-2006}. Because highly reactive chemical species are inevitable by-products of photosynthesis, such regulatory processes are critical for the robustness of photosynthesis. It may be interesting to also investigate such photoprotection mechanisms from the combined multidisciplinary viewpoint of quantum information processing and condensed phase chemical dynamics. 
Recent works by Caycedo-Soler et al. \cite{Cay.Rod.etal-2010}  and by Fassioli and Olaya-Castro \cite{Fas.Ola-2010} are along these lines.

What are the prospects of experimentally verifying the entanglement predicted in the above studies? There are several possible routes to such experimental verification. If one recognizes that in the single excitation subspace entanglement is a result of exciton delocalization, then it is sufficient to measure exciton delocalization lengths in order to verify entanglement. There exist several techniques for measuring exciton delocalization \cite{Mon.Abr.etal-1997, Dah.Pul.etal-2001}: however such measurements have yet to be carried out on the FMO complex or LHCII. In quantum information, a standard method for verifying entanglement is to measure an \textit{entanglement witness}, an observable that is a sufficient condition for entanglement. For light harvesting complexes, within the single excitation approximation, it can be shown that a measure of coherent delocalization provides an entanglement witness \cite{Sar.Ish.etal-2009,Ish.Fle-2010} and this is the most direct route to verifying entanglement in such systems. 
In principle, one could also perform quantum state tomography of the light harvesting complex in order to determine the full density matrix determining the electronic excited state. Recently, techniques for performing tomographic protocols using 
two-color polarization-controlled heterodyne photon-echo experiments have been proposed \cite{Yue.Moh.etal-2010}, although it should be added that the practicality of such proposals is unknown and any tomography of light harvesting complexes is likely to be extremely challenging. Finally, another ambitious route to verifying entanglement is to perform a statistical test that confirms non-classical correlations between spatially separated chromophores -- e.g., measure a Bell inequality \cite{Bel-1964} or CHSH inequality \cite{Cla.Hor.etal-1969} (or even a Leggett-Garg inequality \cite{Wil.McC.etal-2010}). Such measurements would require the ability to measure the electronic excited state of a preferred chromophore in a complex, ideally in a single molecule experiment.  But perhaps more importantly, any such test would require the measurement of two conjugate variables of the electronic excited state; the relevant variables in this context are the presence of an excitation on a chromophore and the relative phase between the ground and excited state of the chromophore. The latter is a particularly challenging measurement since the phase coherence of this optical transition is very short lived, $10 \sim 100\,{\rm fs}$ \cite{Lee.Che.etal-2007,Pre.Ros-1998}.
In summary, while there are in principle several routes to experimentally verifying the predicted entanglement and exploring more general entangled states, all are challenging for current experimental setups.

\bibliographystyle{elsarticle-num}
\bibliography{Whaley_procedia_chem_review_revised.bib}

\begin{thebibliography}{10}
\expandafter\ifx\csname url\endcsname\relax
  \def\url#1{\texttt{#1}}\fi
\expandafter\ifx\csname urlprefix\endcsname\relax\def\urlprefix{URL }\fi
\expandafter\ifx\csname href\endcsname\relax
  \def\href#1#2{#2} \def\path#1{#1}\fi

\bibitem{Ein.Pod.etal-1935}
A.~Einstein, B.~Podolsky, N.~Rosen, Phys. Rev. 47 (1935) 777.

\bibitem{Sch-1935a}
E.~Schr\"odinger, Proc. Cambridge Phil. Soc. 31 (1935) 555.

\bibitem{Bel-1964}
J.S. Bell, Physics 1 (1964) 195.

\bibitem{Cla.Hor.etal-1969}
J.F. Clauser, M.A. Horne, A.~Shimony, R.A. Holt, Phys. Rev. Lett. 23 (1969)
  880.

\bibitem{Leg.Gar-1985}
A.J. Leggett, A.~Garg, Phys. Rev. Lett. 54 (1985) 857.

\bibitem{mikeandike}
M.A. Nielsen, I.L. Chuang, Quantum computation and quantum information,
  Cambridge University Press, 2001.

\bibitem{Ami.Faz.etal-2008}
L.~Amico, R.~Fazio, A.~Osterloh, V.~Vedral, Rev. Mod. Phys. 80 (2008) 517.

\bibitem{Ple.Vir-2007}
M.B. Plenio, S.~Virmani, Quant. Inf. and Comp. 7 (2007) 1.
\newblock \href {http://arxiv.org/abs/0504163 [quant-ph]} {arXiv:0504163
  [quant-ph]}.

\bibitem{Cho-2009}
M.~Cho, Two-dimensional optical spectroscopy, CRC Press, Boca Raton, 2009.

\bibitem{Eng.Cal.etal-2007}
G.S. Engel, T.R. Calhoun, E.L. Read, T.-K. Ahn, T.~Man\v{c}al, Y.-C. Cheng,
  R.E. Blankenship, G.R. Fleming, Nature 446 (2007) 782.

\bibitem{Pan.Hay.etal-2010}
G.~Panitchayangkoon, D.~Hayes, K.A. Fransted, J.R. Caram, E.~Harel, J.~Wen,
  R.E. Blankenship, G.S. Engel, Proc. Natl. Acad. Sci. USA 107 (2010) 12766.
\newblock \href {http://arxiv.org/abs/1001.5108 [physics.bio-ph]}
  {arXiv:1001.5108 [physics.bio-ph]}.

\bibitem{Cal.Gin.etal-2009}
T.R. Calhoun, N.S. Ginsberg, G.S. Schlau-Cohen, Y.-C. Cheng, M.~Ballottari,
  R.~Bassi, G.R. Fleming, J. Phys. Chem. B 113 (2009) 16291.

\bibitem{Col.Won.etal-2010}
E.~Collini, C.Y. Wong, K.E. Wilk, P.M.G. Curmi, P.~Brumer, G.D. Scholes, Nature
  463 (2010) 644.

\bibitem{Lee.Che.etal-2007}
H.~Lee, Y.-C. Cheng, G.R. Fleming, Science 316 (2007) 1462.

\bibitem{Tho.Eck.etal-2009}
M.~Thorwart, J.~Eckel, J.H. Reina, P.~Nalbach, S.~Weiss, Chem. Phys. Lett. 478
  (2009) 234.
\newblock \href {http://arxiv.org/abs/0808.2906 [cond-mat.mes-hall]}
  {arXiv:0808.2906 [cond-mat.mes-hall]}.

\bibitem{Vid.Wer-2002}
G.~Vidal, R.F. Werner, Phys. Rev. A 65 (2002) 032314.

\bibitem{Cai.Pop.etal-2008}
J.~Cai, S.~Popescu, H.J. Briegel (2008).
\newblock \href {http://arxiv.org/abs/0809.4906 [quant-ph]} {arXiv:0809.4906
  [quant-ph]}.

\bibitem{Hil.Woo-1997}
S.~Hill, W.K. Wootters, Phys. Rev. Lett. 78 (1997) 5022.

\bibitem{Bri.Pop-2008}
H.J. Briegel, S.~Popescu (2008).
\newblock \href {http://arxiv.org/abs/0806.4552v2 [quant-ph]}
  {arXiv:0806.4552v2 [quant-ph]}.

\bibitem{Hos.Sch-2010}
H.~Hossein-Nejad, G.D. Scholes, New J. Phys. 12 (2010) 065045.

\bibitem{Ish.Fle-2009}
A.~Ishizaki, G.R. Fleming, J. Chem. Phys. 130 (2009) 234111.

\bibitem{Ish.Cal.etal-2010}
A.~Ishizaki, T.R. Calhoun, G.S. Schlau-Cohen, G.R. Fleming, Phys. Chem. Chem.
  Phys. 12 (2010) 7319.

\bibitem{Sch.Mel.etal-2010}
T.~Scholak, F.~de~Melo, T.~Wellens, F.~Mintert, A.~Buchleitner (2010).
\newblock \href {http://arxiv.org/abs/0912.3560v2 [quant-ph]}
  {arXiv:0912.3560v2 [quant-ph]}.

\bibitem{Reb.Moh.etal-2009}
P.~Rebentrost, M.~Mohseni, I.~Kassal, S.~Lloyd, A.~Aspuru-Guzik, New J. Phys.
  11 (2009) 033003.
\newblock \href {http://arxiv.org/abs/0807.0929 [quant-ph]} {arXiv:0807.0929
  [quant-ph]}.

\bibitem{Ple.Hue-2008}
M.B. Plenio, S.F. Huelga, New J. Phys. 10 (2008) 113019.

\bibitem{Sar.Che.etal-2009}
M.~Sarovar, Y.-C. Cheng, K.B. Whaley (2009).
\newblock \href {http://arxiv.org/abs/0911.5427 [quant-ph]} {arXiv:0911.5427
  [quant-ph]}.

\bibitem{Sar.Ish.etal-2009}
M.~Sarovar, A.~Ishizaki, G.R. Fleming, K.B. Whaley, Nat. Phys. 6 (2010) 462.
\newblock \href {http://arxiv.org/abs/0905.3787 [quant-ph]} {arXiv:0905.3787
  [quant-ph]}.

\bibitem{Ish.Fle-2010}
A.~Ishizaki, G.R. Fleming, New J. Phys. 12 (2010) 055004.

\bibitem{Car.Chi.etal-2010}
F.~Caruso, A.W. Chin, A.~Datta, S.F. Huelga, M.B. Plenio, Phys. Rev. A 81
  (2010) 062346.
\newblock \href {http://arxiv.org/abs/0912.0122 [quant-ph]} {arXiv:0912.0122
  [quant-ph]}.

\bibitem{Fas.Ola-2010}
F.~Fassiolo, A.~Olaya-Castro, New J. Phys. 12 (2010) 085006.
\newblock \href {http://arxiv.org/abs/1003.3610v2 [quant-ph]}
  {arXiv:1003.3610v2 [quant-ph]}.

\bibitem{Moh.Reb.etal-2008}
M.~Mohseni, P.~Rebentrost, S.~Lloyd, A.~Aspuru-Guzik, J. Chem. Phys. 129 (2008)
  174106.
\newblock \href {http://arxiv.org/abs/0805.2741 [quant-ph]} {arXiv:0805.2741
  [quant-ph]}.

\bibitem{Gaa.Bar-2004}
Kevin~M. Gaab, C.J. Bardeen, J. Chem. Phys. 121 (2004) 7813.

\bibitem{Reb.Moh.etal-2009a}
P.~Rebentrost, M.~Mohseni, A.~Aspuru-Guzik, J. Phys. Chem. B 113 (2009) 9942.
\newblock \href {http://arxiv.org/abs/0806.4725 [quant-ph]} {arXiv:0806.4725
  [quant-ph]}.

\bibitem{Fen.Mat-1975}
R.E. Fenna, B.W. Matthews, Nature 258 (1975) 573.

\bibitem{Ame.Val-2000}
H.~van Amerongen, L.~Valkunas, R.~van Grondelle, Photosynthetic Excitons, World
  Scientific, Singapore, 2000.

\bibitem{Bla-2002}
R.E. Blankenship, Molecular mechanisms of photosynthesis, Wiley-Blackwell,
  2002.

\bibitem{Gro.Yu.etal-1998}
M.L. Groot, J.Y. Yu, R.~Agarwal, J.R. Norris, G.R. Fleming, J. Phys. Chem. B
  102 (1998) 5923.

\bibitem{Ved.Ple.etal-1997}
V.~Vedral, M.B. Plenio, M.A. Rippin, P.L. Knight, Phys. Rev. Lett. 78 (1997)
  2275.

\bibitem{Wis.Bar.etal-2004}
H.M. Wiseman, S.D. Bartlett, J.A. Vaccaro, in: Proc. 16th Int. Conf. Laser
  Spec., World Scientific, 2004.

\bibitem{Bar.Rud-2007}
S.D. Bartlett, T.~Rudolph, Rev. Mod. Phys. 79 (2007) 555.

\bibitem{Nit-2006}
A.~Nitzan, Chemical dynamics in condensed phases, Oxford University Press,
  2006.

\bibitem{Joz.Lin-2003}
R.~Jozsa, N.~Linden, Proc. Roy. Soc. London A 459 (2003) 2011.

\bibitem{Hoy.Sar.etal-2009}
S.~Hoyer, M.~Sarovar, K.B. Whaley, New J. Phys. 12 (2010) 065041.
\newblock \href {http://arxiv.org/abs/arXiv:0910.1847 [quant-ph]}
  {arXiv:arXiv:0910.1847 [quant-ph]}.

\bibitem{Car.Chi.etal-2009}
F.~Caruso, A.W. Chin, A.~Datta, S.F. Huelga, M.B. Plenio, J. Chem. Phys. 131
  (2009) 105106.
\newblock \href {http://arxiv.org/abs/0901.4454v2 [quant-ph]}
  {arXiv:0901.4454v2 [quant-ph]}.

\bibitem{Hen.Bel.etal-2009}
E.~Hennebicq, D.~Beljonne, C.~Curutchet, G.D. Scholes, R.J. Silbey, J. Chem.
  Phys. 130 (2009) 214505.

\bibitem{Fas.Naz.etal-2009}
F.~Fassiolo, A.~Nazir, A.~Olaya-Castro, arXiv:0907.5183 [quant-ph] (2009).
\newblock \href {http://arxiv.org/abs/arXiv:0907.5183 [quant-ph]}
  {arXiv:arXiv:0907.5183 [quant-ph]}.

\bibitem{Wu.Liu.etal-2010}
J.~Wu, F.~Liu, Y.~Shen, J.~Cao, R.J. Silbey, arXiv:1008.2236 (2010).
\newblock \href {http://arxiv.org/abs/arXiv:1008.2236} {arXiv:arXiv:1008.2236}.

\bibitem{Moh.Sha.etal-2010}
M.~Mohseni, A.~Shabani, S.~Lloyd, H.~Rabitz, (manuscript in preparation).

\bibitem{Oll.Zur-2001}
H.~Ollivier, W.H. Zurek, Phys. Rev. Lett. 88 (2001) 017901.

\bibitem{Bra.Wil.etal-2009}
K.~Bradler, M.M. Wilde, S.~Vinjanampathy, D.B. Uskov (2009).
\newblock \href {http://arxiv.org/abs/0912.5112 [quant-ph]} {arXiv:0912.5112
  [quant-ph]}.

\bibitem{Cav.Aol.etal-2010}
D.~Cavalcanti, L.~Aolita, S.~Boixo, K.~Modi, M.~Piani, A.~Winter (2010).
\newblock \href {http://arxiv.org/abs/1008.3205 [quant-ph]} {arXiv:1008.3205
  [quant-ph]}.

\bibitem{Mad.Dat-2010}
V.~Madhok, A.~Datta (2010).
\newblock \href {http://arxiv.org/abs/1008.4135 [quant-ph]} {arXiv:1008.4135
  [quant-ph]}.

\bibitem{Ish.Fle-2009a}
A.~Ishizaki, G.R. Fleming, J. Chem. Phys. 130 (2009) 234110.

\bibitem{Ish.Fle-2010:note}
The caption for Fig. 2(b) in Ref. \cite{Ish.Fle-2010} has a typo, where Chl{\it
  a}612 is given instead of Chl{\it b}606.

\bibitem{Bre.Pet-2002}
H.-P. Breuer, F.~Petruccione, The theory of open quantum systems, Oxford
  University Press, 2002.

\bibitem{Hak.Str-1973}
H.~Haken, G.~Strobl, Z. Phys 262~(2) (1973) 135.

\bibitem{Enk-2005}
S.J. van Enk, Single-particle entanglement, Phys. Rev. A 72 (2005) 064306.

\bibitem{Abr.Vor.etal-2008}
D.~Abramavicius, D.V. Voronine, S.~Mukamel, Proc. Natl. Acad. Sci. USA 105
  (2008) 8525.

\bibitem{Kim.Muk.etal-2009}
J.~Kim, S.~Mukamel, G.D. Scholes, Acc. Chem. Res. 42 (2009) 1375.

\bibitem{Muk-2010}
S.~Mukamel, J. Chem. Phys. 132 (2010) 241105.

\bibitem{Dem.Ada.etal-2006}
B.~Demmig-Adams, W.W. Adams, A.~Mattoo, Photoprotection, Photoinhibition, Gene
  Regulation, and Environment, Springer, 2006.

\bibitem{Cay.Rod.etal-2010}
F.~Caycedo-Soler, F.J. Rodr\'iguez, L.~Quiroga, N.F. Johnson, Phys. Rev. Lett.
  104 (2010) 158302.
\newblock \href {http://arxiv.org/abs/1003.2443v1 [cond-mat.soft]}
  {arXiv:1003.2443v1 [cond-mat.soft]}.

\bibitem{Mon.Abr.etal-1997}
R.~Monshouwer, M.~Abrahamsson, F.~van Mourik, R.~van Grondelle, J. Phys. Chem.
  B 101 (1997) 7241.

\bibitem{Dah.Pul.etal-2001}
M.~Dahlbom, T.~Pullerits, S.~Mukamel, V.~Sundstr\"om, J. Phys. Chem. B 105
  (2001) 5515.

\bibitem{Yue.Moh.etal-2010}
J.~Yuen-Zhou, M.~Mohseni, A.~Aspuru-Guzik (2010).
\newblock \href {http://arxiv.org/abs/1006.4866 [quant-ph]} {arXiv:1006.4866
  [quant-ph]}.

\bibitem{Wil.McC.etal-2010}
M.M. Wilde, J.M. McCracken, A.~Mizel, Proc. R. Soc. A 466 (2010) 1347.
\newblock \href {http://arxiv.org/abs/0911.1097 [quant-ph]} {arXiv:0911.1097
  [quant-ph]}.

\bibitem{Pre.Ros-1998}
O.V. Prezhdo, P.J. Rossky, Phys. Rev. Lett. 81 (1998) 5294.

\end{thebibliography}

\end{document}